\documentclass[A4,11pt]{article}
\usepackage{cite}
\usepackage{graphicx}
\usepackage{epstopdf}
\usepackage{amsmath}
\usepackage{amssymb}
\usepackage{amsfonts}
\usepackage{amsthm}
\usepackage{mathrsfs}
\usepackage{ccaption}
\usepackage{comment}
\usepackage{overpic}
\usepackage{array}
\usepackage{epsfig}

\newcommand{\al}{\alpha}
\newcommand{\be}{\beta}
\newcommand{\de}{\delta}

\newcommand{\nn}{\nonumber}

\def\N{{\cal N}}

\def\so{\mathfrak{so}}
\def\su{\mathfrak{su}}
\def\sp{\mathfrak{sp}}
\def\sl{\mathfrak{sl}}

\def\osp{\mathfrak{osp}}
\def\C{\mathbb{C}}

\newcommand{\dlb}{\ensuremath{[\![}}
\newcommand{\drb}{\ensuremath{]\!]}}

\newcommand{\la}{\langle}
\newcommand{\ra}{\rangle}

\def\al{\alpha}
\def\be{\beta}
\def\ga{\gamma}
\def\de{\delta}

\def\cG{\mathcal{G}}

\def\cN{\mathcal{N}}

\def\nn{\nonumber}

\usepackage[
      colorlinks=true,
      linkcolor=blue,
      urlcolor=blue,
      filecolor=blue,
      citecolor=red,
      pdfstartview=FitV,
      pdftitle={},
      pdfauthor={},
      pdfsubject={},
      pdfkeywords={},
      pdfpagemode=None,
      bookmarksopen=true
]{hyperref}

\numberwithin{equation}{section}

\begin{document}

\hfill
{\tt ULB-TH/10-33}

\vspace{1cm}

\begin{center}
{\Large \bf $\mathcal{N}=5$ three-algebras and $5$-graded Lie superalgebras
}\\[20mm]
{\bf Sung-Soo Kim and  Jakob Palmkvist}  \\[5mm]
{\it Physique Th\'{e}orique et Math\'{e}matique\\
Universit\'{e} Libre de Bruxelles\\ and\\ International Solvay Institutes,\\
ULB-C.P. 231, B-1050 Bruxelles,
Belgium}\\[7mm]
{\tt sungsoo.kim@ulb.ac.be, jakob.palmkvist@ulb.ac.be}\\[7mm]

\end{center}

\vspace{8mm}

\begin{abstract}
We discuss a generalization of $\N=6$ three-algebras to $\mathcal{N}=5$ three-algebras in connection to anti-Lie triple systems and basic Lie superalgebras of type II. We then show that the structure constants defined in anti-Lie triple systems agree with those of $\N=5$ superconformal theories in three dimensions.
\end{abstract}

\newpage

\newpage
%%%%%%%%%%%%%%%%%%%%%%%%%%%%%%%%%%%%
\setcounter{equation}{0}
%%%%%%%%%%%%%%%%%%%%%%%%%%%%%%%%%%%%%%%%%%
\section{Introduction and summary}
Since the pioneering work by Bagger and Lambert \cite{Bagger:2007jr}, and Gustavsson \cite{Gustavsson:2007vu} (BLG), where three-algebras were used to construct the ${\cal N}=8$ superconformal theory in three dimensions, such structures have played an important r\^{o}le toward a better understanding of M-theory. 
This $\N=8$ theory allows only an SO(4) gauge symmetry. More general gauge groups, $\mathrm{SU}(n)\times \mathrm{SU}(n)$ and 
${\rm U}(n) \times {\rm U}(n)$,  were soon considered by 
Aharony, Bergman, Jafferis and Maldacena (ABJM) \cite{Aharony:2008ug} in theories with $\N=6$ supersymmetries. The corresponding three-algebra construction was found in \cite{Bagger:2008se}.

In \cite{Hosomichi:2008jb} 
Hosomichi, Lee, Lee, Lee and Park constructed a 
superconformal three-dimensional theory with $\mathcal{N}=5$ supersymmetry 
and $\mathrm{Sp}(2n)\times \mathrm{O}(m)$ gauge symmetry. (Similar theories were constructed in \cite{Aharony:2008gk}.) It was described by embedding the corresponding Lie algebra into the Lie superalgebra $B(m,\,n)= \osp(2m+1|2n)$ or  $D(m,\,n)=\osp(2m|2n)$. When the same construction is applied to the Lie superalgebras $A(m,\,n)=\sl (m+1|n+1)$ and $C(n+1)=\osp (2|2n)$, the supersymmetry enhances from $\mathcal{N}=5$ to $\mathcal{N}=6$ \cite{Hosomichi:2008jb}. In the case of $A(m,\,n)$ this is the ABJM theory. Additional $\mathcal{N}=5$ theories, based on the exceptional Lie superalgebras, 
$F(4),\,G(3)$ and $D(2,1;\alpha)$, were found in \cite{Bergshoeff:2008bh} by use of the embedding tensor approach. 

The connection between Lie superalgebras and superconformal theories 
was first noticed in \cite{Gaiotto:2008sd}. The odd part determines the representations for matter fields and the even part corresponds to the gauge group. It turned out that two important types of Lie superalgebras, called basic Lie superalgebras of type I and type II, are relevant for theories with $\cN=6$ and $\cN=5$ supersymmetry, respectively.
Indeed, among the Lie superalgebras mentioned above, $A(m,\,n)$ and $C(n+1)$ are the basic ones of type I, whereas $B(m,\,n)$, $D(m,\,n)$,  $F(4)$, $G(3)$ and $D(2,1;\,\alpha)$, are those of type II. The connection between Lie superalgebras and superconformal theories has also been discussed
 in \cite{Chen:2010xj}.

It was shown in \cite{Palmkvist:2009qq} that basic Lie superalgebras of type I are in one-to-one correspondence with simple 
$\mathcal{N}=6$ three-algebras. 
These two algebraic structures were related by generalized Jordan triple systems, which were also considered in \cite{Nilsson:2008kq}. 
In an alternative approach \cite{deMedeiros:2008zh} 
it was shown in \cite{FigueroaO'Farrill:2009pa} that basic Lie superalgebras are in one-to-one correspondence with simple 
quaternionic anti-Lie triple systems. The most general $\cN=5$ theory was constructed from such a triple system in 
\cite{deMedeiros:2009eq} 
(in $\cN=1$ superfield language), and from a 
`symplectic three-algebra' in \cite{Chen:2009cwa}.
Studying $\cN=5$ theories based on an arbitrary triple system, Bagger and Bruhn determined in
\cite{Bagger:2010zq} the conditions that this triple system must satisfy. The resulting conditions are the same as those defining a 
`symplectic three-algebra' in \cite{Chen:2009cwa}. Furthermore, this is indeed an anti-Lie triple system.

In this note, we generalize the $\cN=6$ three-algebras defined in \cite{Palmkvist:2009qq} to $\cN=5$ three-algebras. We show that the $\cN=5$ theories 
can be formulated in terms of anti-Lie triple systems which are 
obtained from basic Lie superalgebras and thus
related to $\cN=5$ three-algebras.
We explain how the $\cN=5$ three-algebras in turn lead to $\N=6, 8$ three-algebras by imposing further conditions in the definition.
After reviewing the construction in \cite{Bagger:2010zq} of $\cN=5$ theories, we study the possible representations of the structure constants in section \ref{sec:repre} and show that the representations agree with \cite{Bergshoeff:2008bh}.

%%%%%%%%%%%%%%%%%%%%%%%%%%%%%%%%%%%%%%%%%%%
\section{Graded Lie superalgebras}\label{GLS}

In this section we review some facts from the theory of Lie superalgebras. For details, we refer to \cite{Kac77A,Frappat}. Any Lie superalgebra $\mathcal{G}$ 
can be written as a direct sum of two subspaces $\mathcal{G}_{(0)}$ and $\mathcal{G}_{(1)}$ (the even and odd part) such that
\begin{align} \label{z2grading}
\dlb \mathcal{G}_{(p)},   \mathcal{G}_{(q)}\drb \subseteq  \mathcal{G}_{(p+q)},
\end{align}
where the subscripts are counted mod 2. 
A Lie superalgebra $\mathcal{G}$ may also be written as a sum of subspaces $\mathcal{G}_k$
for any integer $k$, such that $\mathcal{G}_k\subset\mathcal{G}_{(0)}$ if $k$ is even and $\mathcal{G}_k\subset\mathcal{G}_{(1)}$
if $k$ is odd. Then $\mathcal{G}$ is said to have a consistent $\mathbb{Z}$-grading. When we henceforth talk about a 3-graded or 5-graded Lie superalgebra we refer to a consistent $\mathbb{Z}$-grading such that $\mathcal{G}_k=0$ for $|k|\geq 2$ or $|k|\geq 3$, respectively. 

It follows from (\ref{z2grading}) that $\mathcal{G}_{(0)}$ is a subalgebra (which is an ordinary Lie algebra) and that 
$\mathcal{G}_{(1)}$ is a representation of $\mathcal{G}_{(0)}$. The Lie superalgebra $\mathcal{G}$ is said to be 
{\it classical} if this representation is completely reducible. 
Then there are two cases, which divide the classical Lie superalgebras into two types: type I and type II. 
The representation of $\mathcal{G}_{(0)}$ on $\mathcal{G}_{(1)}$ is either
a direct sum of two irreducible representations (type I), or
irreducible (type II).
A classical Lie superalgebra $\mathcal{G}$ is said to be {\it basic} if 
it admits a non-degenerate bilinear form $\kappa$ that is {\it invariant}, which means
\begin{align}
\kappa(\dlb x,\,y\drb,\,z)=\kappa(x,\,\dlb y,\,z\drb)
\end{align}
for all $x,\,y,\,z \in \mathcal{G}$. This bilinear form will furthermore satisfy
\begin{align}
\kappa(x,\,y)&=\tfrac12\big((-1)^p+(-1)^q\big)\,\kappa(y,\,x),
\end{align}
for all $x\in \mathcal{G}_{(p)}$ and $y\in \mathcal{G}_{(q)}$. Thus it satisfies the requirements for an {\it inner product} \cite{Frappat}. We will occasionally write $\kappa(x,\,y)=\la x | y \ra$.

Any basic Lie superalgebra admits a 3-grading if it is of type I, and a 5-grading if it is of type II. The inner product is such that $\kappa(x,\,y)$, where 
$x \in \cG_i$ and $y \in \cG_j$, is nonzero only if $i+j=0$.
There is also an antilinear map $\tau : \mathcal{G}_k \to \mathcal{G}_{-k}$ such that $\tau^2(x)=(-1)^px$ for any $x \in \mathcal{G}_{(p)}$, and
\begin{align}
\dlb \tau(x),\,\tau(y)\drb = \tau(\dlb x,\,y\drb)
\end{align} 
for any $x,\,y\in \mathcal{G}$. We call such a map $\tau$ a {\it graded superconjugation}. The antilinearity of $\tau$ means that
$\tau(\alpha x)=\alpha^\ast \tau(x)$ if $\alpha^\ast$ is the conjugate of a complex number $\alpha$.

Let $\mathcal{G}$ be a basic Lie superalgebra with inner product $\kappa$ and graded superconjugation $\tau$. Let $M^m$ be a basis for $\mathcal{G}_{(0)}$ and $Q^a$ a basis for $\mathcal{G}_{(1)}$.
Then we write
\begin{align}
k^{mn}&=k^{nm}=\kappa(M^m,\,M^n), & \omega^{ab}=
-\omega^{ba}&=\kappa(Q^a,\,Q^b).
\end{align}
Let $k_{mn}$ and $\omega_{ab}$ be the inverses of 
$k^{mn}$ and $\omega^{ab}$,
\begin{align}
k^{mp}k_{pn}&=\delta^m{}_n &
\omega^{ac}\omega_{cb}&=-\delta^a{}_b.
\end{align}
We use these tensors to raise and lower indices (with the convention $X^a=\omega^{ab}X_b$ and $X_a=-\omega_{ab}X^b$).
Now there are structure constants $(t_m)^{{a}{b}}$ and $f^{mn}{}_p$ such that\begin{align} 
[M^m,\,M^n]&=f^{mn}{}_p M^p, & [M^m,\,Q^a]&=(t^m)^{a}{}_b Q^b, &\label{commMQ}
\{Q^a,\,Q^b \}&=(t_m)^{ab}M^m. &
\end{align}
We will use these structure constants in the next section to construct the structure constants 
of an anti-Lie triple system.

%%%%%%%%%%%%%%%%%%%%%%%%%%%%%%%%%%%%%%%%%%%%%
\section{Three-algebras and triple systems} \label{three-alg-section}
With a triple system we here simply mean a complex vector space $V$ with a triple product  $f : V \times V \times V \to V$ that is linear or antilinear in each argument.  
By imposing further conditions, one obtains different kinds of triple systems, some of which are called `three-algebras'.

The original notion of a three-algebra \cite{Bagger:2007jr} was generalized in \cite{Bagger:2008se}. In \cite{Palmkvist:2009qq} these triple systems were called $\mathcal{N}=8$ three-algebras and $\mathcal{N}=6$ three-algebras, respectively. We will follow the terminology in this note, but also generalize the notion further to triple systems that we call $\mathcal{N}=5$ three-algebras.
%%%%%%%%%%%%%%%%%%%%%%%%%%%%%%%%%%%%%%%%
\subsection{Three-algebras} \label{three-alg-subsection}
An $\mathcal{N}=5$ { three-algebra} is a triple system $V$ with a triple product $f: V \times V \times V \to V$  and an `inner product' $h: V \times V \to \C$, such that
\begin{enumerate}
\item the triple product $(xyz) \equiv f(x,\,y,\,z)$ 
is linear in $x$ and $z$ but antilinear in $y$:
\begin{align}
\al (xyz)=((\al x)yz)=(x(\al^\ast y)z)=(xy(\al z))
\end{align}
for any complex number $\alpha$ (where $\ast$ is the complex conjugate),
\item the triple product satisfies 
\begin{align}
(uv(xyz))&=((uvx)yz)-(x(vuy)z)+(xy(uvz)),\label{N=5ta1}\\
K_{xy}(K_{uv}(z))&=(K_{xy}(v)uz)+(K_{xy}(u)vz)
\label{N=5ta2},
\end{align}
where $K_{xy}(z)=(xzy)+(yzx)$,
\item the inner product $\la x ,\, y \ra \equiv h(x,\,y)$ is linear in $x$ and antilinear in $y$,
\item the inner product satisfies 
\begin{align}
\la w ,\, (xyz)\ra=\la y ,\, (zwx) \ra&=\la(wzy),\,x\ra=\la(yxw),\,z\ra, 
\label{N=5ta3}\\
\la x ,\,y\ra&=\la y ,\,x\ra^{\ast}, \label{N=5ta4}
\end{align}
\item the inner product is positive-definite.
\end{enumerate}

By imposing further conditions one obtains $\cN=6$ and $\cN=8$ three-algebras.
An $\mathcal{N}=6$ { three-algebra} is an $\mathcal{N}=5$ three-algebra with $K_{xy}=0$ for any $x,\,y$. This means that the triple product is antisymmetric in the first and third arguments,
\begin{align}
(xzy)=-(yzx).
\end{align}
Thus (\ref{N=5ta2}) is trivially satisfied.
An $\mathcal{N}=8$ { three-algebra} is an $\mathcal{N}=6$ three-algebra $V$ with a \textit{conjugation} $C$ (an antilinear involution) such that 
the triple product satisfies $(xC(y)z)=-(yC(x)z)$ and the inner product $h$ is real. This implies that the triple product is totally antisymmetric and that the inner product is symmetric.

%%%%%%%%%%%%%%%%%%%%%%%%%%%%%%
\subsection{Anti-Lie triple systems}\label{subsec:LTS}
An anti-Lie triple system
is a triple system with a triple product 
$[xyz]$ that is trilinear and
satisfies 
\begin{align}
[uv[xyz]]-[xy[uvz]]&=[[uvx]yz]+[x[uvy]z],\label{alts1}\\
[xyz]&=[yxz],\label{alts2}\\
[xyz]+[yzx]+[zxy]&=0.
\label{alts3}
\end{align}
With the opposite sign in (\ref{alts2}) we would get a Lie triple system instead.

The anti-Lie triple systems in \cite{Chen:2009cwa,Bagger:2010zq}, 
furthermore, have a bilinear form such that
\begin{align} \label{bilinform}
\la w ,\, [xyz] \ra &= \la y ,\, [zwx] \ra,\nn\\
\la x ,\,y \ra&=-\la y ,\,x \ra.
\end{align}
This is also true for the quaternionic anti-Lie triple systems considered in \cite{FigueroaO'Farrill:2009pa}, but these have in addition a `quaternionic' structure, which is a vector space automorphism $J$ such that $J^2=-1$ and
\begin{align}
[J(x)J(y)J(z)]=J([xyz]).
\end{align}

%%%%%%%%%%%%%%%%%%%%%%%%%%%%%%%
\subsection{Connection to Lie superalgebras}

For any Lie superalgebra $\cG$, the odd subspace $\cG_{(1)}$ is a triple system under the triple product
\begin{align} \label{alts-product}
[XYZ]=[\{X,Y\},Z],
\end{align}
where $X,Y,Z \in \mathcal{G}_{(1)}$.
The general properties that such a triple product satisfies (by the Jacobi superidentity and the symmetries of the superbracket) are exactly those that define an anti-Lie triple system (in the same way as an ordinary Lie algebra leads to a Lie triple system). 
The structure constants can be obtained from \eqref{commMQ},
\begin{align}
g^{abcd} \equiv
\kappa([\{Q^a,\,Q^b \}, Q^c], Q^d)=
(t_m){}^{ab} (t^m){}^{cd}.
\end{align}
Conversely, any anti-Lie triple system gives rise to a Lie superalgebra $\mathcal{G}$ \cite{2010arXiv1010.3599C}.

In the case of a basic Lie superalgebra $\mathcal{G}$ with a 3- or 5-grading, the graded superconjugation $\tau$ becomes a quaternionic structure $J$ on the anti-Lie triple system $\mathcal{G}_{(1)}$. 
We can use $\tau$ to
decompose each element $X \in \mathcal{G}_{(1)}$ into a sum 
$X=x+\tau(y)$,
where $x,y \in \mathcal{G}_{-1}$.
Since $\mathcal{G}$ is either 3- or 5-graded, we have
\begin{align}
[\{x,y\},z]=[\{\tau(x),\tau(y)\},\tau(z)]=0
\end{align}
for any $x,y,z \in \mathcal{G}_{-1}$. Then, by use of the Jacobi superidentity, any triple product
(\ref{alts-product}),
where $X,Y,Z \in \mathcal{G}_{(1)}$, decomposes into a sum of triple products
\begin{align} \label{three-alg-triple-product}
(xyz)=[\{x,\tau(y)\},z],
\end{align}
where $x,y,z \in \mathcal{G}_{-1}$. 
It is straightforward to verify that $\mathcal{G}_{-1}$ now satisfies the definition above of an 
$\cN=5$ three-algebra with the triple product
(\ref{three-alg-triple-product}) and the inner product
\begin{align}
h(x,y)=\kappa(x,\tau(y)).
\end{align}
Furthermore, when $\mathcal{G}$ is 3-graded, the $\cN=5$ three-algebra reduces to an $\cN=6$ three-algebra.
We have thus shown that an anti-Lie triple system obtained from a basic Lie superalgebra gives rise to an $\cN=5$ three-algebra.

In this note we show that the anti-Lie triple systems that have been used in $\cN=5$ theories can be obtained from basic Lie superalgebras of type II. Thus they give rise to $\cN=5$ three-algebras that could be used instead of anti-Lie triple systems.
Since the $\cN=5$ three-algebras, unlike anti-Lie triple systems, are generalizations of the $\cN=6$ three-algebras, they are (in our opinion) more natural to use in the construction of $\cN=5$ theories. These could then be unified with the $\cN=6$ theories in a way similar to the approach in \cite{Chen:2009cwa}. Such a construction was performed
in \cite{Palmkvist:2011aw}
showing explicitly that $\cN=5$ three-algebras indeed lead to $\cN=5$ theories.

%%%%%%%%%%%%%%%%%%%%%%%%%%%%%%%%%%%%%%%%%%%%%%

\section{Construction of the $\mathcal{N}=5$ theory based on anti-Lie triple systems}\label{sec:N=5}

Here we review the construction on $\mathcal{N}=5$ theories by Bagger and Bruhn \cite{Bagger:2010zq}.
Keeping global symmetry to be Sp(4) R-symmetry,
they found non-trivial supersymmetry transformations\footnote{It is not difficult to see that the supersymmetry transformations \eqref{newsusy} are equivalent to those in \cite{Hosomichi:2008jb}
by introducing an embedding tensor
$\frac{4\pi}{3}(t_m)^{ab}(t^m)^{cd}=f^{abcd}$.}
\begin{align}
\delta Z^{Ad}=&~i \bar \xi^{AD}\Psi^d_{D},\cr
\delta \Psi_{D}^d=&~\xi_{AD}\,\slash\!\!\!\!{D}Z^{Ad} +\frac{1}{2}g^{bcad}Z^A_aZ^B_bZ^C_c\xi_{DC}\omega_{AB}
+g^{acbd}Z^A_aZ^B_bZ^C_c\xi_{AB}\omega_{DC},\cr
\delta {A}_\mu{}^a{}_d=&~
\frac{3i}{2}g^{bca}{}_d\omega^{BE}\bar\xi_{EC}\gamma_\mu\Psi_{Bb}Z^C_c, \label{newsusy}
\end{align}
where  $Z^{Ad}$, $\Psi_{D}^d$ are the scalars and fermions, respectively,  transforming in the bifundamental representation of the gauge group, and ${A}_\mu{}^a{}_d$ denote the gauge fields.
The capital letters $A,B,\ldots$ are for $\sp(4)$ R-symmetry indices, and lower case letters $a, b,\ldots$ for gauge group indices. The $\sp(4)$ invariant tensor $\omega_{AB}$ satisfying $\omega^{AB}\omega_{BC}=-\delta^A{}_C$ is used to raise and lower the indices, e.g.,
$X^A=\omega^{AB}X_B$ and $X_A=-\omega_{AB}X^B$. In the same way, the gauge indices are raised and lowered by an invariant tensor $J^{ab}J_{bc}=-\delta^a{}_c$. The structure constants
 \begin{align}
g^{abcd}=g^{bacd}=g^{cdab}\label{gsymm}
\end{align}
satisfy the cyclicity condition
\begin{align}
g^{(abc)d}=0,\label{BBcyclicity}
\end{align}
and the fundamental identity
\begin{align}\label{BBFI}
g^{abhe}g_e{}^{fcd}+
g^{abfe}g_e{}^{hcd}+
g^{abce}g_e{}^{dhf}+
g^{abde}g_e{}^{chf}=0,
\end{align}
which is of the same form as that for ${\cal N}=6$ theories.

The supersymmetry algebras close on a translation and a gauge transformation. For instance, the supersymmetry transformations on the scalars are given by
\begin{align}
[\delta_1,\delta_2] Z^A_d = \frac{i}{2}\bar\xi_{2}^{BC}\gamma^\mu\xi_{1\,BC
}\,D_\mu Z^A_d+\delta_{\Lambda}Z^A_d,
\end{align}
where $\xi_{BC}$ are real antisymmetric supersymmetry transformation parameters and the gauge transformation is given by
\begin{align}\label{gaugetransBB}
\delta_{\Lambda}Z^A_d=\Lambda^a{}_d Z^A_a=-
 \frac{3i}{4}\bar\xi_{[2}^{DF}\xi_{1]\,BF}Z^{B}_bZ^C_c\omega_{DC}g^{bca}{}_dZ^A_a.
\end{align}

The representations of $g^{abcd}$ are characterized by $\so(m)$ and $\sp(2n)$ algebras. Then the combinations that satisfy \eqref{BBcyclicity} and \eqref{BBFI} are
\begin{align}
g^{aibjckdl}&=(\de^{ac}\de^{bd}-\de^{ad}\de^{bc})J^{ij}J^{kl}
-  (J^{ik}J^{jl}+J^{jk}J^{il})\de^{ab}\de^{cd}, \label{g1g2}\\
g^{aibjckdl}&=\de^{ac}\de^{bd} J^{ik}J^{jl}
+\de^{ad}\de^{bc}J^{jk}J^{il},\label{g3}
\end{align}
where $a,b,\ldots$ denote the indices for $\so(m)$ and $i,j,\ldots$ for $\sp(2n)$.
The first combination \eqref{g1g2} leads to
$\sp(2n)\oplus \so(m)$ transformations 
\begin{align}
\delta Z^{Adl}=-\frac{3i}{2} \bar\xi_{[2}^{DF}\xi_{1]\,BF}\omega_{DC}
\Big(Z^B_{bk}Z^{Cl}_b Z^{Adk}
+Z^{Bk}_b Z^{Cd}_{k}Z^{Al}_{b}\Big).
\end{align}
The second combination \eqref{g3} leads to 
 $\sp(2mn)$ transformations
\begin{align}
\delta Z^{Adl} &=3i \bar\xi_{[2}^{DF}\xi_{1]\,BF}\omega_{DC}
Z^B_{bk}Z^{Cdl}Z^{Ak}_b= \Lambda_{bk}{}^{dl}Z^{Ak}_b,\label{sptrans}
\end{align}
since $\Lambda^{bk\,dl} = \Lambda^{dl\,bk}$.

We introduce a map \cite{Bagger:2008se} $g$: $\sp(N)\rightarrow \sp(N)$
\begin{align}
g(\lambda)^a{}_d = \lambda_{bc}\, g^{bca}{}_d ,
\end{align}
it then follows from the fundamental identity that
\begin{align}
[g(\lambda_1), g(\lambda_2)] = g(\lambda_3),\label{matrixcomm}
\end{align}
where $\lambda_{3\,bc}=-f(\lambda_1)^{e}{}_c\lambda_{2\,be}  -f(\lambda_1)^{e}{}_b\lambda_{2\,ce}$. This means that the gauge transformations act as a matrix commutator, implying that they indeed form a Lie subalgebra.
In what follows, we examine how these $\N=5$ theories can be understood from the anti-Lie triple systems.

\section{Connection to anti-Lie triple systems}\label{sec:repre}
We now relate the three-algebra constructions to the anti-Lie triple system that we discussed in section \ref{subsec:LTS}. First we introduce the basis of the Lie algebra of the gauge transformations such that the fields take the form of $Z^A = Z^A_a T^a$. Then the anti-Lie triple product is given by 
\begin{align}
[T^a\,T^b\,T^c]= - g^{abc}{}_d\, T^d ,
\end{align}
where, by construction, the first two indices are symmetric $g^{abc}{}_d= g^{bac}{}_d$. The structure constants are given by
\begin{align}
g^{abcd} = \la [T^aT^bT^c]|T^d\ra.
\end{align}
The identity \eqref{alts1} implies that 
 the gauge transformation acts as a derivation
\begin{align}
\delta[Z^A\,Z^B\,Z^C] = [\delta Z^A\,Z^B\,Z^C]+[Z^A\,\delta Z^B\,Z^C]+[Z^A\,Z^B\,\delta Z^C].
\end{align}
It is straightforward to see that \eqref{alts1} is equivalent to the fundamental identity \eqref{BBFI}.

\subsection{Basic Lie superalgebras of type II}

The Lie superalgebras $B(m,\,n)$ and $D(m,\,n)$, are the algebras $\osp(m|2n)$ for $m=1$ and $m=3,\,4,\ldots$. When $m=2$ we have instead $\osp(2|2n)=C(n+1)$, which is a basic Lie algebra of 
type I, and thus associated with an $\mathcal{N}=6$ three-algebra, see \cite{Palmkvist:2009qq}. 

As a first example we consider $\osp(1|2N)$. Its even subalgebra is $\sp(2N)$, spanned by $N(2N+1)$ symmetric 
generators $M^{IJ}$, and the odd subspace is spanned by $2N$ generators $Q^I$,
where $I, J=1,\ldots, 2N$.
The commutation relations read
\begin{align}
\{ Q^I,\, Q^J \} &= M^{IJ},\nn\\
[ M^{IJ},\, Q^K ] &= \Omega^{JK}Q^I + \Omega^{IK}Q^J,\nn\\
[ M^{IJ},\, M^{KL} ] &=\Omega^{JK}M^{IL}+\Omega^{IK}M^{JL}+\Omega^{JL}M^{IK}+\Omega^{IL}M^{JK}, \label{kommrelosp22m}
\end{align}
where
$\Omega^{IJ}$ is the antisymmetric invariant tensor of $\sp(2N)$ satisfying 
\begin{align}
\Omega^{IJ}\Omega_{JK} =-\delta^I{}_K. 
\end{align}
The inner product is given by 
$\la Q^I|Q^J\ra=\Omega^{IJ}$.
Using this, we find that the structure constants of the
anti-Lie triple system are
\begin{align} \label{osp12nalt}
\la [Q^{I}Q^{J}Q^K] | Q^L \ra &= \la [\{Q^I,\,Q^J\},\,Q^K] |Q^L \ra\nn\\
&=\la \Omega^{JK}Q^I+\Omega^{IK}Q^J | Q^L\ra = \Omega^{JK}\Omega^{IL}+\Omega^{IK}\Omega^{JL}.
\end{align}

Suppose that $N = mn$ for some integers $m,\,n\geq 1$. 
We can then decompose each
$\sp(2N)$ index into one $\so(m)$ index 
($a,\,b,\,\ldots=1,\,2,\,\ldots,\,m$) and one 
$\sp(2n)$ index ($i,\,j,\,\ldots=1,\,2,\,\ldots,\,2n$), so that
\begin{align}
\Omega^{IJ}=\Omega^{ai\,bj}=\delta^{ab}J^{ij},
\end{align}
where $\delta^{ab}$ and $J^{ij}$ are the invariant tensors of the corresponding $\so(m)$ and $\sp(2n)$ subalgebras.
Under this decomposition, the structure constants \eqref{osp12nalt} become
\begin{align}
\la [Q^{ai}Q^{bj}Q^{ck}] | Q^{dl} \ra &= \Omega^{bj\,ck}\Omega^{ai\,dl}+\Omega^{ai\,ck}\Omega^{bj\,dl}\nn\\
&= \de^{bc}\de^{ad}J^{jk}J^{il}+\de^{ac}\de^{bd}J^{ik}J^{jl},
\label{strucsp2mn}
\end{align}
which are the structure constants \eqref{g3} leading to the $\sp(2mn)$ gauge transformations.
Now we turn to $\osp(m|2n)$ for $m=3,\,4,\,\ldots$ and $n=1,\,2,\,\ldots$. 
The even subalgebra is $\so(m) \oplus \sp(2n)$, spanned by the antisymmetric generators $M^{ab}$ and the symmetric generators 
$M^{ij}$. The odd subspace is spanned by $2mn$ elements $Q^{ai}$.
The 
commutation relations read
\begin{align}
[M^{ab},M^{cd}] &= \delta^{bc}M^{ad} -\delta^{bd}M^{ac} -\delta^{ac}M^{bd}+\delta^{ad}M^{bc},\nn\\
[M^{ij},M^{kl}] &= J^{jk}M^{il} +J^{jl}M^{ik} +J^{ik}M^{jl}+J^{il}M^{jk},\nn\\
[M^{ab},Q^{c i}] &= \delta^{bc}Q^{a i}-\delta^{ac}Q^{b i},\nn\\
[M^{ij},Q^{a k}] &= J^{jk}Q^{a i}+J^{ik}Q^{a j},\nn\\
\{Q^{a i},Q^{b j}\}&= v(J^{ij}M^{ab}+\delta^{ab}M^{ij}), \label{kommrelospn2m}
\end{align}
where $v$ is a normalization constant. The inner product is given by 
$\la Q^{ai}|Q^{bj}\ra=\delta^{ab}J^{ij}$, and
the structure constants of the
anti-Lie triple system are
then
\begin{align} \label{osp12naltssc}
\la [Q^{ai}Q^{bj}Q^{ck}] | Q^{dl} \ra &= \la [\{Q^{ai},\,Q^{bj}\},\,Q^{ck}] |Q^{dl} \ra\nn\\
&=v\la [J^{ij}M^{ab}+\delta^{ab}M^{ij},\,Q^{ck}]| Q^{dl} \ra \ra\nn\\
&=v\big(J^{ij}J^{kl}(\de^{bc} \de^{ad}-\de^{ac}\de^{bd})+\de^{ab} \de^{cd}( J^{jk}J^{il}+J^{ik}J^{jl})\big).
\end{align}
For $v=-1$, this agrees with \eqref{g1g2}, yielding the $\so(m)\oplus\sp(2n)$ gauge transformations.

\subsection{Exceptional Lie superalgebras}
The construction of anti-Lie triple systems from Lie superalgebras
can be applied to the exceptional cases $F(4)$, $G(3)$, and $D(2,1;\al)$. We end this note by showing that the construction reproduces the structure constants discussed in \cite{Bergshoeff:2008bh} as well as \cite{Bagger:2010zq}.

\subsubsection{$F(4)$}
The even part of the Lie superalgebra $F(4)$ is $\su(2)\oplus\so(7)$ spanned by
the $\su(2)$ generators $S^i$ ($i=1,2,3$) and the $\so(7)$ generators $M^{ab}$ ($a,b=1,\ldots,7$). The odd part is spanned by
$Q_{\al m}$ where $\al=+1,-1$ and $m=1,\ldots,8$.
The commutation relations \cite{Frappat,Nahm-scheunert} are 
\begin{align}
[S^i,S^j]&= i\, \epsilon^{ijk}S^k,& [S^i,M^{ab}]&=0,\nn\\
[S^i, Q_{\al m}]&=\frac{1}{2}\sigma^i_{\be\al}Q_{\be m},&
[M^{ab}, Q_{\al m}]&=\frac{1}{2}(\Gamma^a \Gamma^b )_{nm}Q_{\al n},\nn
\end{align}
\vspace{-.5cm}
\begin{align}
[M^{ab},M^{cd}] &= \delta^{bc}M^{ad} -\delta^{bd}M^{ac} -\delta^{ac}M^{bd}+\delta^{ad}M^{bc},\cr
\{Q_{\al m},Q_{\be n}\}&=v\, 2 \tilde C_{mn} (C^{(2)}\sigma^j)_{\al\be}S^j +v  \frac13 C^{(2)}_{\al\be}(\tilde C\,\Gamma^a\Gamma^b)_{mn}M^{ab},\,
\end{align}
where $v$ is a normalization constant,  $\sigma^i$ are the Pauli matrices, $C^{(2)}=i\sigma^2$, $\tilde  C$ is the $8\times 8$ charge conjugation matrix with
\begin{align}
(\tilde  C)^T = \tilde  C,\qquad (\Gamma^a)^T = - \tilde  C\, \Gamma^a, 
\end{align}
and $\Gamma^a$ are 8-dimensional gamma matrices satisfying 
$\{\Gamma^a,\Gamma^b\} = 2\delta^{ab}$.
The inner product is then
$\la Q_{\al m} | Q_{\be n} \ra = \epsilon_{\al\be}\tilde C_{mn}$
and the structure constants are
\begin{align}
g_{m\al\, n\be \,p\gamma\, q\de}&=\la [\{Q_{\al m},Q_{\be n}\}, Q_{\gamma p}]|Q_{\de_q}\ra \cr
&\!\!\!\!\!\!\!\!\!\!\!\!\!\!\!\!\!\!\!\!\!\!\!\!=
v \Big( \tilde C_{mn} \tilde C_{pq} (\epsilon_{\al\gamma}\epsilon_{\be\de}-\epsilon_{\gamma\be}\epsilon_{\al\de}) +\frac16 \epsilon_{\al\be}\epsilon_{\gamma\de}
(\tilde C\Gamma^a\Gamma^b)_{mn}(\tilde C\Gamma^a\Gamma^b)_{qp}\Big),
\end{align}
where we used the completeness relations of the Pauli matrices 
\begin{align}
(\sigma^i)_{\al\be}(\sigma^i)_{\gamma\delta} = 2 \delta_{\al\delta}\delta_{\beta\gamma} - \delta_{\al\be}\delta_{\gamma\delta},
\end{align}
and the cyclicity condition 
\begin{align}
\epsilon_{\al\be}\epsilon_{\ga\de}+\epsilon_{\be\ga}\epsilon_{\al\de}+\epsilon_{\ga\al}\epsilon_{\be\de}=0.\label{su2cyclic}
\end{align}
One may take $\tilde C_{mn} = \delta_{mn}$ and $v=-1/2$ to obtain
\begin{align}
g_{m\al\, n\be \,p\gamma\, q\de}&=
\frac12 \de_{mn} \de_{pq} (\epsilon_{\gamma\al}\epsilon_{\be\de} +
\epsilon_{\gamma\be}\epsilon_{\al\de}) +
\frac{1}{12} \epsilon_{\al\be}\epsilon_{\gamma\de}
(\Gamma^a\Gamma^b)_{mn}(\Gamma^a\Gamma^b)_{pq}.
\end{align}

%%%%%%%%%%%%%%%%%%%%%%%%%%%%%%%

\subsubsection{$G(3)$}

The even part of the Lie superalgebra $G(3)$ is $\su(2)\oplus G_2$ spanned by the $\su(2)$ generators $S^i$ ($i=1,2,3$) and the $G_2$ generators $M^{ab}$ ($a,b=1,\ldots,7$) obeying $\xi^{abc}M^{ab}=0$ \cite{Frappat,Nahm-scheunert}, where $\xi^{abc}$ is a totally anti-symmetric $G_2$ invariant tensor whose nonvanishing components are
\begin{align}
\xi^{123}=\xi^{145}=\xi^{176}=\xi^{246}=\xi^{257}=\xi^{347}=\xi^{365}=1.
\end{align}
The odd part is spanned by $Q^{\al a}$ ($\al=+1,-1$).  
The commutations relations are
\begin{align}
[S^i,S^j]&= i\, \epsilon^{ijk}S^k,& [S^i,M^{ab}]&=0,&
[S^i, Q^{\al a}]&=\frac{1}{2}\sigma^i{}^{\be\al}Q^{\be a},\nn
\end{align}
\vspace{-1cm}
\begin{align}
[M^{ab},M^{cd}] &= 3 \delta^{bc}M^{ad} -3\delta^{bd}M^{ac} -3\delta^{ac}M^{bd}+3\delta^{ad}M^{bc}- \xi^{abe}\xi^{cdf}M^{ef},\nn\\
[M^{ab}, Q^{\al c}]&=2 \delta^{ac}Q^{\al b}-2 \delta^{bc}Q^{\al a}-\eta^{abcd}Q^{\al d},\nn\\
\{Q^{\al m},Q^{\be n}\}&=2 v\delta^{ab}(C^{(2)}\sigma^j)^{\al\be}S^j -\frac{v}{2}C^{(2)}{}^{\al\be}M^{mn},
\end{align}
where $C^{(2)}=i\sigma^2$, and  $\eta^{abcd}$ is a totally antisymmetric tensor whose nonvanishing components are
\begin{align}
\eta^{1247}=\eta^{1265}=\eta^{1364}=\eta^{1375}=\eta^{2345}=\eta^{2376}=\eta^{4576}=1,
\end{align}
satisfying
\begin{align}
\eta^{abcd}=\de^{ad}\de^{bc}-\de^{ac}\de^{bd}+ \xi^{abe}\xi^{cde}.
\end{align}
The inner product is defined as 
$\la Q^{\al a} | Q^{\be b} \ra = \epsilon^{\al\be}\de^{ab}$,
and the structure constants are
\begin{align}
g^{\al a\, \be b\,\gamma c\, \de d}&=\la [\{Q^{\al b},Q^{\be b} \}, Q^{\gamma d} ]|Q^{\de d} \ra \cr
&\!\!\!\!\!\!\!\!\!\!\!\!\!\!\!\!\!\!\!\!\!\!\!\!=v \Big( 
\delta^{ab}\delta^{cd}(\epsilon^{\al\gamma}\epsilon^{\be\delta}+\epsilon^{\be\gamma}\epsilon^{\al\delta}) -\epsilon^{\al\be}\epsilon^{\gamma\delta}(
\de^{ac}\de^{bd}-\de^{ad}\de^{bc}-\frac{1}{2}\eta^{abcd}
)
\Big).\quad
\end{align}

%%%%%%%%%%%%%%%%%%%%%%%%%%%%%%%

\subsubsection{$D(2,1;\al)$}
The even part of the Lie superalgebra $D(2,1;\al)$ is 
$\so(4)\oplus \sp(2)$
spanned by
the antisymmetric generators $M^{ab}$ ($a,b=1,\ldots,4$) and the symmetric generators 
$M^{ij}$ ($i=1,2$). The odd part is spanned by generators
$Q^{ai}$. It has a free parameter $\alpha \neq -1,\,0$ \cite{Frappat}. 
The commutation relations are of the same form as \eqref{kommrelospn2m} (with $\epsilon^{ij}$ instead of $J^{ij}$) except for
\begin{align}
\{Q^{a i},Q^{b j}\}&= v\Big(\epsilon^{ij}(M^{ab}+ \frac{\beta}{2} \varepsilon^{abcd}M^{cd})+\delta^{ab}M^{ij}\Big)
\end{align}
where $\varepsilon^{abcd}$ is the totally antisymmetric invariant tensor of $\so(4)$
and $\beta$ is a free parameter. Comparing with the construction in \cite{Frappat} we find that the relation between $\alpha$ and $\beta$ is
\begin{align}
\frac{\beta}2 = \frac{1-\alpha}{1+\alpha}.
\end{align}
It follows from the inner product 
$\la Q^{ai} | Q^{ bj} \ra = \epsilon^{ij}\de^{ab}$
that
\begin{align}
&g^{aibjckdl}
=\la [\{Q^{ai},Q^{bj}\},Q^{ck}] | Q^{dl} \ra \cr
&=v\Big((\de^{ad}\de^{bc}-\de^{ac}\de^{bd}-\beta\,\varepsilon^{abcd}\,\big) \epsilon^{ij}\epsilon^{kl}+ (\epsilon^{ik}\epsilon^{jl}+\epsilon^{jk}\epsilon^{il})\de^{ab}\de^{cd}\Big).
\end{align}

\section*{Acknowledgments}
We thank Jos\'e Figueroa-O'Farrill, Kimyeong Lee, Carlo Maccaferri, Carlo Meneghelli, Bengt E.W.~Nilsson and Constantinos Papageorgakis for discussions.
We also thank Jonathan Bagger for correspondence and for sharing the new version of \cite{Bagger:2010zq} before it had appeared. Finally we thank the referee for providing  constructive comments and pointing out typos. 
Our work is supported by IISN - Belgium (conventions 4.4511.06 and 4.4514.08), by the Belgian Federal Science Policy Office 
through the Interuniversity Attraction Pole P6/11.

%%%%%%%%%%%%%%%%%%%%%%%%%%%%%%%%%%%%%%
%%\bibliographystyle{utphysmod2}
%\bibliographystyle{utphys}
%%\bibliography{biblio}
%\begin{thebibliography}{100}
%\end{thebibliography}
%\newpage

\providecommand{\href}[2]{#2}\begingroup\raggedright\endgroup
\end{document}